\documentclass[journal]{IEEEtran} 				
\IEEEoverridecommandlockouts 						
\usepackage{amsmath,amssymb}
\usepackage{amsthm}
\usepackage{float}
\usepackage{color}
\usepackage[mathscr]{eucal}
\usepackage{graphics,graphicx,multicol}
\usepackage{epstopdf}
\usepackage{enumerate}
\usepackage{subfigure}
\usepackage{algorithmic}
\usepackage{algorithm}
\usepackage{morefloats}
\usepackage{enumerate}
\usepackage{subfigure}
\usepackage{multirow}
\usepackage{array}
\usepackage{pstricks, pst-node, pst-plot, pst-circ}
\usepackage{moredefs}
\usepackage{courier}
\usepackage{mathtools}
\usepackage{amsbsy}

\begin{document}
\title{A Price Selective Centralized Algorithm for Resource Allocation with Carrier Aggregation in LTE Cellular Networks}
\author{Haya Shajaiah, Ahmed Abdelhadi and Charles Clancy \\
Bradley Department of Electrical and Computer Engineering\\
Hume Center, Virginia Tech, Arlington, VA, 22203, USA\\
\{hayajs, aabdelhadi, tcc\}@vt.edu\\
}
\maketitle
\begin{abstract}
In this paper, we consider a resource allocation with carrier aggregation optimization problem in long term evolution (LTE) cellular networks. In our proposed model, users are running elastic or inelastic traffic. Each user equipment (UE) is assigned an application utility function based on the type of its application. 
Our objective is to allocate multiple carriers resources optimally among users in their coverage area while giving the user the ability to select one of the carriers to be its primary carrier and the others to be its secondary carriers. The UE's decision is based on the carrier price per unit bandwidth. 
We present a price selective centralized resource allocation with carrier aggregation algorithm to allocate multiple carriers resources optimally among users while providing a minimum price for the allocated resources. In addition, we analyze the convergence of the algorithm with different carriers rates. Finally, we present simulation results for the performance of the proposed algorithm.
\end{abstract}
\begin{keywords}
Resource Allocation with Carrier Aggregation, Elastic Traffic, Inelastic Traffic
\end{keywords}
\providelength{\AxesLineWidth}       \setlength{\AxesLineWidth}{0.5pt}%
\providelength{\plotwidth}           \setlength{\plotwidth}{8cm}
\providelength{\LineWidth}           \setlength{\LineWidth}{0.7pt}%
\providelength{\MarkerSize}          \setlength{\MarkerSize}{3pt}%
\newrgbcolor{GridColor}{0.8 0.8 0.8}%
\newrgbcolor{GridColor2}{0.5 0.5 0.5}%
\section{Introduction}\label{sec:intro}
In recent years, the number of mobile subscribers and their traffic have increased rapidly. Network providers are now offering multiple services such as multimedia telephony and mobile-TV \cite{QoScontrol}. More spectrum is required to meet these demands. 
Therefore, allowing mobile users to employ multiple carriers by aggregating their different frequency bands is needed \cite{UtilityMax}. The carrier aggregation (CA) feature was added to the 3GPP LTE standard in Release 10 \cite{Yuan10CarrierAggregation}. 
This feature allows the LTE Advanced to meet the International Mobile Telecommunications (IMT) requirements for the fourth-generation standards defined by the International Telecommunications Union (ITU) \cite{Evolution}. A resource allocation with carrier aggregation optimization problem is formulated in \cite{Haya_Utility1}. The authors proposed two-stage distributed resource allocation (RA) algorithm that allocates the primary and secondary carriers resources optimally among users.

In this paper, we formulate the RA with CA problem into a convex optimization framework. We use logarithmic utility functions to represent delay-tolerant applications and sigmoidal-like utility functions to represent real-time applications running on the UEs subscribing for a mobile service. The primary and secondary carriers optimization problems assign part of the bandwidth from the multiple carriers to each user. A minimum QoS is guaranteed for each user 
by using a proportional fairness approach. Our centralized RA with CA algorithm provides a minimum price per unit bandwidth by allowing users under the coverage area of multiple evolved node Bs (eNodeB)s to select the carrier with the lowest price to be their primary carrier and the others to be their secondary carriers. This mechanism allows users to improve their allocated rates by using the CA feature while maintaining the lowest possible price for their allocated aggregated rates. Additionally, our centralized algorithm is performed mostly in the eNodeBs which reduces the transmission overhead created by the distributed algorithm introduced in \cite{Haya_Utility1}. 
\subsection{Related Work}\label{sec:related}
In \cite{kelly98ratecontrol}, the authors introduced bandwidth proportional fair resource allocation with logarithmic utilities. The algorithms at the links are based on Lagrange multiplier methods of optimization theory. 
In \cite{Fundamental}, the authors used sigmoidal-like utility functions to represent real-time applications. In \cite{RebeccaThesis}, the authors proposed weighted aggregated utility functions for the elastic and inelastic traffic. 
An optimal resource allocation algorithm is presented in \cite{Ahmed_Utility1} and \cite{Ahmed_Utility2} to allocate a single carrier resources optimally among mobile users. 
In \cite{Ahmed_Utility3}, two-stage resource allocation algorithm is proposed to allocate the eNodeB resources among users running multiple applications at a time. 
In \cite{Haya_Utility2}, a resource allocation optimization problem is presented for two groups of users. The two groups are public safety users group and commercial users group. 
In \cite{Haya_Utility4}, the authors presented a resource allocation with users discrimination algorithms to allocate the eNodeB resources optimally among users and their applications. A resource allocation optimization problem with carrier aggregation is presented in \cite{Haya_Utility3} to allocate resources from the LTE Advanced carrier and the MIMO radar carrier to each UE in a LTE Advanced cell based on the running application of the UE.
\subsection{Our Contributions}\label{sec:contributions}
Our contributions in this paper are summarized as:
\begin{itemize}
\item We present a resource allocation optimization problem with carrier aggregation that solves for logarithmic and sigmoidal-like utility functions.
\item We propose a price selective centralized RA with CA algorithm to allocate multiple carriers resources optimally among users.
\item We show that our algorithm is a robust one that converges to the optimal rates whether the eNodeBs available resources are abundant or scarce. We present simulation results for the performance of our resource allocation algorithm.

\end{itemize}

The remainder of this paper is organized as follows. Section \ref{sec:problemFormulation} presents the problem formulation. In Section \ref{sec:rate allocation}, we present our RA with CA optimization problem. 
In section \ref{sec:algorithm}, we present our centralized rate allocation with CA algorithm for the utility proportional fairness optimization problem. In section \ref{sec:sim}, we discuss simulation setup and provide quantitative results along with discussion. Section \ref{sec:conclude} concludes the paper.
\section{Problem Formulation}\label{sec:problemFormulation}
\begin{figure}[H]
\centering
\includegraphics[height=2.4in, width=2.1in]{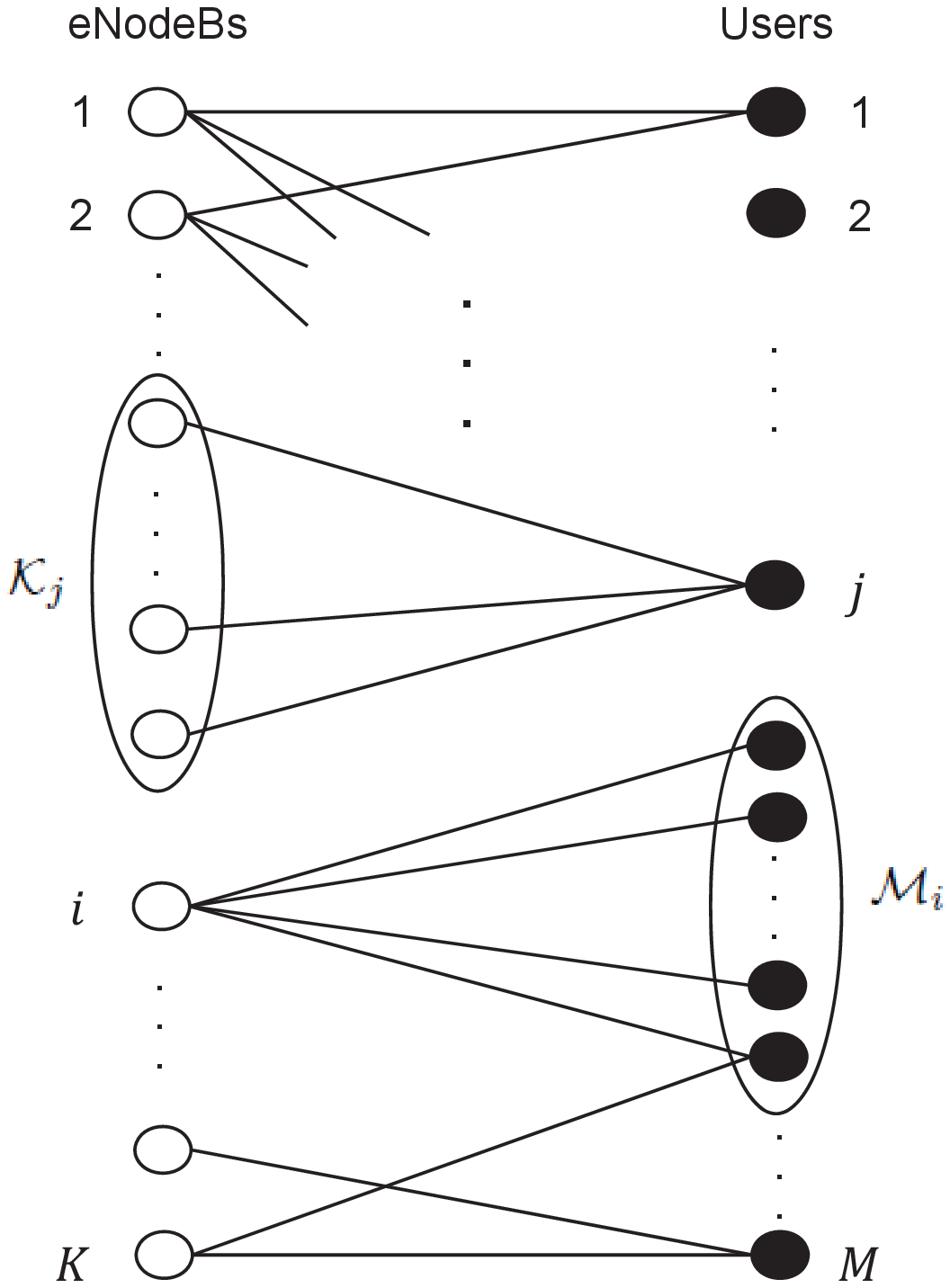}
\caption{System model for a LTE mobile system with $M$ users and $K$ carriers eNodeBs. $\mathcal{M}_i$ represents the set of users located under the coverage area of the $i^{th}$ eNodeB and $\mathcal{K}_j$ represents the set of all in range eNodeBs for the $j^{th}$ user.}
\label{fig:System__Model}
\end{figure}

We consider a LTE mobile system with $M$ users and $K$ carriers eNodeBs, one eNodeB in each cell, as illustrated in Figure \ref{fig:System__Model}. The users located under the coverage area of the $i^{th}$ eNodeB are forming a set of users $\mathcal{M}_i$ where $\mathcal{M}_i \in \{\mathcal{M}_1,\mathcal{M}_2,...,\mathcal{M}_{K}\}$ and $M_i=|\mathcal{M}_i|$ is the number of users in the users set $\mathcal{M}_i$ under the coverage area of the $i^{th}$ eNodeB. Each joint user $j$ is located under the coverage area of a set of eNodeBs, as shown in Figure \ref{fig:System__Model}, that is given by $\mathcal{K}_j$ where $\mathcal{K}_j \in \{\mathcal{K}_1,\mathcal{K}_2,...,\mathcal{K}_{M}\}$ and $K_j=|\mathcal{K}_j|$ is the number of eNodeBs in the set $\mathcal{K}_j$ of all in range eNodeBs for user $j$.

Each eNodeB calculates its offered price per unit bandwidth (assuming it is the primary carrier for all users under its coverage area) and provides each user under its coverage area with its offered price. Each joint user selects the carrier with the least offered price to be its primary carrier and the rest of all in range carriers to be its secondary carriers. The eNodeB with the least offered price first allocates its resources to all users under its coverage area based on the applications running on their UEs. The remaining eNodeBs then start allocating their resources in the order of their offered prices to all users under their coverage area based on the users applications and the rates that are allocated to the joint users from other eNodeBs (with lower offered prices).

We express the user satisfaction with its provided service using utility functions 
\cite{DL_PowerAllocation} \cite{Fundamental} \cite{Utility-proportional}. We assume that the $j^{th}$ user' application utility function $U_j (r_j)$ is strictly concave or sigmoidal-like function where $r_j$ is the rate allocated to the $j^{th}$ user. Delay tolerant applications are represented by logarithmic utility functions whereas real-time applications are represented by sigmoidal-like utility functions. These utility functions have the following properties:
\begin{itemize}
\item $U_j(0) = 0$ and $U_j(r_j)$ is an increasing function of $r_j$.
\item $U_j(r_j)$ is twice continuously differentiable in $r_j$ and bounded above.
\end{itemize}

We use the normalized sigmoidal-like utility function in our model, same as the one presented in \cite{DL_PowerAllocation}, that is
\begin{equation}\label{eqn:sigmoid}
U_j(r_j) = c_j\Big(\frac{1}{1+e^{-a_j(r_j-b_j)}}-d_j\Big)
\end{equation}
where $c_j = \frac{1+e^{a_jb_j}}{e^{a_jb_j}}$ and $d_j = \frac{1}{1+e^{a_jb_j}}$ so it satisfies $U_j(0)=0$ and $U_j(\infty)=1$. The normalized sigmoidal-like function has an inflection point at $r_j^{\text{inf}}=b_j$. Additionally, we use the normalized logarithmic utility function, used in \cite{Ahmed_Utility1}, that can be expressed as
\begin{equation}\label{eqn:log}
U_j(r_j) = \frac{\log(1+k_jr_j)}{\log(1+k_jr_{\text{max}})}
\end{equation}
where $r_{\text{max}}$ gives $100\%$ utilization and $k_j$ is the slope of the curve that varies based on the user application. So, it satisfies $U_j(0)=0$ and $U_j(r_{\text{max}})=1$.
\section{Multiple Carriers Optimization Problem}\label{sec:rate allocation}
In this section we formulate the RA problem for allocating the primary and secondary carriers resources optimally among users under their coverage areas. Each carrier first calculates its offered price per unit bandwidth assuming that it is the primary carrier for all UEs under its coverage area. Then, each carrier starts allocating its available resources optimally among all users in its coverage area in the order of the carrier's offered price, such that the carrier with a lower offered price perform the RA prior to the one with a higher offered price.
\subsection{The Price Selection Problem and enodeB Sorting}\label{subsec:primary_carrier}
As mentioned earlier, each carrier calculates its offered price assuming it is the primary carrier for all users under its coverage area. The carrier's offered price is obtained from the following RA optimization problem:
\begin{equation}\label{eqn:opt1_prob_fairness}
\begin{aligned}
& \underset{\mathbf{r_i}}{\text{max}}
& & \prod_{j=1}^{M_i}U_j(r_{i,j}) \\
& \text{subject to}
& & \sum_{j=1}^{M_i}r_{i,j} \leq R_i,\\
& & & r_{i,j} \geq 0, \;\;\;\;\; j = 1,2, ...,M_i.\\
\end{aligned}
\end{equation}
where $\mathbf{r_i} =\{r_{i,1},r_{i,2},...,r_{i,M_i}\}$, $M_i$ is the number of UEs under the coverage area of the $i^{th}$ eNodeB and $R_i$ is the maximum achievable rate of the $i^{th}$ eNodeB. The resource allocation objective function is to maximize the total system utility when allocating the eNodeB resources. Furthermore, it provides proportional fairness among utilities. Therefore, no user is allocated zero resources and a minimum QoS is provided to each user. Real-time applications are given priory when allocating the eNodeB resources using this approach. Optimization problem (\ref{eqn:opt1_prob_fairness}) is a convex optimization problem and there exists a unique tractable global optimal solution \cite{Ahmed_Utility1}. The objective function in optimization problem (\ref{eqn:opt1_prob_fairness}) is equivalent to $\underset{\mathbf{r_i}}{\text{max}} \sum_{j=1}^{M_i} \log U_j(r_{i,j})$.

From optimization problem (\ref{eqn:opt1_prob_fairness}), we have the Lagrangian:
\begin{equation}\label{eqn:Lagrangian Primary 1}
\begin{aligned}
L_i(r_{i,j}) = &(\sum_{j=1}^{M_i}\log U_j(r_{i,j}))\\
&-p_i^{\text{offered}}(\sum_{j=1}^{M_i}r_{i,j}-R_i-z_i)
\end{aligned}
\end{equation}
where $z_i\geq 0$ is the slack variable and $p_i^{\text{offered}}$ is the Lagrange multiplier which is equivalent to the shadow price that corresponds to the $i^{th}$ carrier price per unit bandwidth for the $M_i$ channels as in \cite{Ahmed_Utility1}. The set of all carriers in the LTE mobile system is given by $\mathcal{K}=\{1,2,...,K\}$ and their corresponding offered prices are given by $\mathcal{P}^{\text{offered}}=\{p_1^{\text{offered}},p_2^{\text{offered}},...,p_K^{\text{offered}}\}$. The $j^{th}$ user set of all in range carriers $\mathcal{K}_j$ (i.e. $\mathcal{K}_j=\{1,2,...,K_j\}$) corresponding offered prices are given by $\mathcal{P}^j=\{p_1^j,p_2^j,...,p_{K_j}^j\}$.

All in range carriers $\mathcal{K}_j$ of the $j^{th}$ user are arranged based on their offered prices as follows:
\begin{align*}
l_1^j=
& \arg \underset{\mathcal{K}_j}  \min \{p_1^j,p_2^j,...,p_{K_j}^j\} \\
l_2^j=
& \arg \underset{\mathcal{K}_j-\{l_1^j\}} \min \{p_1^j,p_2^j,...,p_{K_j}^j\} \\
&\vdots \\
l_{K_j}^j=
& \arg \underset{\mathcal{K}_j-\{l_1^j,...,l_{K_j-1}^j\}} \min \{p_1^j,p_2^j,...,p_{K_j}^j\}\\
\end{align*}
where $l_1^j$ is the carrier with the lowest offered price and $l_{K_j}^j$ is the carrier with the highest offered price within the $j^{th}$ user set $\mathcal{K}_j$ of all in range carriers and $\mathcal{P}^j=\{p_1^j,p_2^j,...,p_{K_j}^j\}$ is the set of the offered prices of all in range carriers for the $j^{th}$ user. The $j^{th}$ user sends an assignment of $1$ to the $i^{th}$ eNodeB that is corresponding to eNodeB $l_1^j$ (i.e. the eNodeB with the least offered price among the $j^{th}$ user's all in range carriers). On the other hand, the $j^{th}$ user sends an assignment of $0$ to each of the remaining eNodeBs in its range. Once the $i^{th}$ eNodeB receives an assignment of $1$ from each UE in its coverage area it starts allocating its resources to the $M_i$ UEs in $\mathcal{M}_i$ such that the $j^{th}$ UE is allocated an optimal rate $r_i^{j,opt}$ from the $i^{th}$ eNodeB. Once the $j^{th}$ UE is allocated rate from its primary carrier $l_1^j$, it then sends an assignment of $1$ to the $i^{th}$ eNodeB that is corresponding to eNodeB $l_2^j$ and sends an assignment of $0$ to each of the remaining eNodeBs in its range. The process continues until the $j^{th}$ UE sends an assignment of $1$ to the $i^{th}$ eNodeB that is corresponding to eNodeB $l_{K_j}^j$ and receives its allocated rate from that eNodeB. The $j^{th}$ UE then calculates its aggregated final optimal rate $r_j^{agg}$.
%
\subsection{RA Optimization Problem}\label{subsec:secondary_carrier}
Once the carriers offered prices are calculated as discussed in \ref{subsec:primary_carrier}, each user $j$ selects eNodeB $l_1^j$ to be its primary carrier and the remaining carriers in its range to be its secondary carriers. The eNodeB with the least offered price is the first one to start allocating its resources among all users in its coverage area. Each of the remaining eNodeBs then starts allocating its available resources after all the users in its coverage area are allocated rates from carriers in their range with lower offered prices.
Eventually, each user $j$ is allocated rates from all of the $K_j$ carriers in its range. As discussed before, the $i^{th}$ carrier eNodeB starts allocating its resources among all users in its coverage area once it receives an assignment of $1$ from each of the $M_i$ users in $\mathcal{M}_i$. The rate allocated to the $j^{th}$ user from its $i^{th}$ carrier is given by $r_i^{j,opt}$.

The RA optimization problem for the $i^{th}$ carrier eNodeB in $\mathcal{K}$, such that the $i^{th}$ eNodeB received an assignment of $1$ from each of the users under its coverage area, 
can be written as:
\begin{equation}\label{eqn:opt2_prob_fairness}
\begin{aligned}
& \underset{\mathbf{r_i}}{\text{max}}
& & \prod_{j=1}^{M_i}U_j(r_i^j+c_i^j)\\
& \text{subject to}
& & \sum_{j=1}^{M_i}r_i^j \leq R_i,\\
& & &  r_i^j \geq 0, \;\;\;\;\; j = 1,2, ...,M_i,\\
& & &  c_i^j = \sum_{n=1, n \neq i}^{K}v_n^j r_n^{j,opt},\\
& & &  v_n^j =  \left\{
    \begin{array}{ll}
      1 , \;\; \text{the $j^{th}$ UE $\in$ $\mathcal{M}_n,$} \\
      0 , \;\; \text{the $j^{th}$ UE $\notin$ $\mathcal{M}_n,$} \\
    \end{array}
  \right.
\end{aligned}
\end{equation}
where $\mathbf{r_i} =\{r_i^1,r_i^2,...,r_i^{M_i}\}$, $R_i$ is the $i^{th}$ eNodeB available resources, $c_i^j$ is equivalent to the total rates allocated to the $j^{th}$ user by the carriers in its range with lower offered prices than the $i^{th}$ carrier offered price, $v_n^j$ is equivalent to $1$ if the $j^{th}$ UE $\in$ $\mathcal{M}_n$ and is equivalent to $0$ if the $j^{th}$ UE $\notin$ $\mathcal{M}_n$
%
and $r_n^{j,opt}$ is the optimal rate allocated to the $j^{th}$ user by the $n^{th}$ eNodeB (i.e. the $n^{th}$ carrier $\in \mathcal{K}$). 
Once the $j^{th}$ user is allocated rate from all the carriers in its range, it then calculates its aggregated final optimal rate $r_j^{agg}=\sum_{i=1}^{K}v_i^j r_i^{j,opt}$.

Optimization problem (\ref{eqn:opt2_prob_fairness}) gives priority to the real-time application users and ensures that the minimum rate allocated to each user is $c_i^j$. Optimization problem (\ref{eqn:opt2_prob_fairness}) is a convex optimization problem and there exists a unique tractable global optimal solution \cite{Ahmed_Utility1}. The objective function in optimization problem (\ref{eqn:opt2_prob_fairness}) is equivalent to $\underset{\mathbf{r_i}}{\text{max}} \sum_{j=1}^{M_i} \log U_j(r_i^j+c_i^j)$.

From optimization problem (\ref{eqn:opt2_prob_fairness}), we have the Lagrangian:
\begin{equation}\label{eqn:Lagrangian Secondary 1}
\begin{aligned}
L_i(r_i^j) = &(\sum_{j=1}^{M_i}\log U_j(r_i^j+c_i^j))\\
&-p_i(\sum_{j=1}^{M_i}r_i^j-R_i-z^i)
\end{aligned}
\end{equation}
where $z^i\geq 0$ is the slack variable and $p_i$ is the Lagrange multiplier which is equivalent to the shadow price that corresponds to the $i^{th}$ carrier price per unit bandwidth for the $M_i$ channels as in \cite{Ahmed_Utility1}.
\section{Algorithm }\label{sec:algorithm}
In this section, we present our price selective centralized RA with CA algorithm. Each UE is allocated optimal rates from its all in range carriers and the final optimal rate allocated to each UE is the aggregated rate. The algorithm starts when each UE transmits its application parameters to all in range eNodeBs. Each eNodeB assigns initial values $w_{i,j}(0)$ to the users applications. Each eNodeB performs an internal iterative algorithm to calculate its offered price per unit bandwidth. In each iteration, the eNodeB checks the difference between the current value $w_{i,j}(n)$ and the previous one $w_{i,j}(n-1)$, if the difference is greater than a threshold $\delta$, the shadow price $p_i^{\text{offered}}(n)=\frac{\sum_{j=1}^{M_i}w_{i,j}(n)}{R_i}$ is calculated by the eNodeB. 
Each eNodeB uses $p_i^{\text{offered}}(n)$ to calculate the rate $r_{i,j}(n)$ that is the solution of the optimization problem $r_{i,j}(n)=\arg \underset{r_{i,j}}\max \Big(\log U_j(r_{i,j})-p_i^{\text{offered}}(n)r_{i,j}\Big)$. The calculated rate is then used to calculate a new value $w_{i,j}(n)$ where $w_{i,j}(n)=p_i^{\text{offered}}(n) r_{i,j}(n)$. Each eNodeB checks the fluctuation condition as in \cite{Ahmed_Utility2} and calculates a new value $w_{i,j}(n)$. Once the difference between the current $w_{i,j}(n)$ and the previous one is less than $\delta$ for all UEs, the $i^{th}$ eNodeB sends its offered price $p_i^{\text{offered}}$ to all UEs in its coverage area.
%
%
\begin{algorithm}
\caption{The $j^{th}$ UE Algorithm} \label{alg:UE_FK}
\begin{algorithmic}
\STATE {Let $c_i^j = 0\:\:\forall i \in \{1,2,...,K\}$}%
\STATE {Send the UE application utility parameters $k_j$, $a_j$ and $b_j$ to all in range eNodeBs}%
\STATE {Receive offered prices that are equivalent to $\mathcal{P}^j=\{p_1^j,p_2^j,...,p_{K_j}^j\}$ from all in range carriers eNodeBs}%
\LOOP
    \FOR{$m \leftarrow 1$  to  $K_j$}%
\STATE {$l_m^j=\arg \underset{\mathcal{K}_j-\{l_1^j,...,l_{m-1}^j\}} \min \{p_1^j,p_2^j,...,p_{K_j}^j\}$ is carrier $l_m^j$ for the $j^{th}$ UE}%
    \ENDFOR
\ENDLOOP
\LOOP
    \FOR{$m \leftarrow 1$  to  $K_j-1$}%
\STATE {Send Flag assignment of $1$ to the $i^{th}$ eNodeB and an assignment of $0$ to the remaining carriers in $\mathcal{K}_j$ \COMMENT{eNodeB $i$ $=$ eNodeB $l_m^j$}}%
\STATE {Send $c_i^j$ to the $i^{th}$ eNodeB \COMMENT{eNodeB $i$ $=$ eNodeB $l_m^j$}}%
\STATE {Receive the optimal rate $r_i^{j,opt}$ from the $i^{th}$ eNodeB \COMMENT{eNodeB $i$ $=$ eNodeB $l_m^j$}}%
\STATE {Receive shadow price $p_i$ from the $i^{th}$ eNodeB \COMMENT{eNodeB $i$ $=$ eNodeB $l_m^j$}}%
\STATE {Send the optimal rate $r_i^{j,opt}$ to the $i^{th}$ eNodeB \COMMENT{the $i^{th}$ eNodeB corresponds to eNodeB $l_{m+1}^j$}}%
\STATE {Calculate new $c_i^j=\sum_{n=1, n \neq i}^{K}v_n^j r_n^{j,opt}$ for the $i^{th}$ eNodeB that corresponds to eNodeB $l_{m+1}^j$}%
    \ENDFOR
\ENDLOOP
\STATE {Send $c_i^j$ to the $i^{th}$ eNodeB \COMMENT{the $i^{th}$ carrier corresponds to carrier $l_{K_j}^j$}}%
\STATE {Receive the optimal rate $r_i^{j,opt}$ from the $i^{th}$ eNodeB \COMMENT{the $i^{th}$ carrier corresponds to carrier $l_{K_j}^j$}}%
\STATE {Receive shadow price $p_i$ from the $i^{th}$ eNodeB \COMMENT{the $i^{th}$ carrier corresponds to carrier $l_{K_j}^j$}}%
\STATE {Calculate the aggregated final optimal rate $r_j^{agg}=c_i^j+r_i^{j,opt}$ \COMMENT{the $i^{th}$ carrier corresponds to carrier $l_{K_j}^j$}}%
\end{algorithmic}
\end{algorithm}

Once the $j^{th}$ UE receives the offered prices $p_i^{\text{offered}}$ from all in range carriers, it sends an assignment of $1$ to the $i^{th}$ eNodeB with the lowest offered price that is corresponding to eNodeB $l_1^j$ and an assignment of $0$ to the remaining eNodeBs in its range. The $j^{th}$ UE then receives its allocated rate $r_i^{j,opt}$ and shadow price $p_i$ from that eNodeB. It then updates the $c_i^j$ value and sends it to the $i^{th}$ eNodeB that is corresponding to eNodeB $l_2^j$, it also sends an assignment of $1$ to that eNodeB and an assignment of $0$ to the remaining eNodeBs in its range. The process continues until the $j^{th}$ UE receives its allocated rate $r_i^{j,opt}$ and shadow price $p_{K_j}$, it then calculates its aggregated final optimal rate $r_j^{agg}$.

On the other hand, Once the $i^{th}$ eNodeB receives assignments of $1$ from all UEs in its coverage area it calculates the optimal rate $r_i^{j,opt}$ and shadow price $p_i$ and sends them to each UE in its coverage area. The process continues until the eNodeB with the highest offered price receives assignment of $1$ from all UEs in its coverage area, it then sends each of these UEs its allocated optimal rate $r_i^{j,opt}$ and shadow price $p_i$.

\begin{algorithm}
\caption{The $i^{th}$ eNodeB Algorithm} \label{alg:eNodeB_FK}
\begin{algorithmic}
	\STATE{Let $w_{i,j}(0) = 0\:\:\forall j \in \mathcal{M}_i$}
    \STATE {Receive application utility parameters $k_j$, $a_j$ and $b_j$ from all UEs under the coverage area of the $i^{th}$ eNodeB}%
\LOOP
    \WHILE {$|w_{i,j}(n) -w_{i,j}(n-1)|>\delta$ for any $j=\{1,....,M_i\}$ where the $j^{th}$ UE under the coverage area of the $i^{th}$ eNodeB} %
    \STATE {Calculate $p_i^{\text{offered}}(n) = \frac{\sum_{j=1}^{M_i}w_{i,j}(n)}{R_i}$}
    \FOR{$j \leftarrow 1$  to  $M_i$}%
    \STATE {Solve $r_{i,j}(n) = \arg \underset{r_{i,j}}\max \Big(\log U_j(r_{i,j}) - p_i^{\text{offered}}(n)r_{i,j}(n)\Big)$}%
    \STATE {Calculate new $w_{i,j}(n)= p_i^{\text{offered}}(n) r_{i,j}(n)$}
        \IF {$|w_{i,j}(n) - w_{i,j}(n-1)| > \Delta w$}
    \STATE {$w_{i,j}(n) =w_{i,j}(n-1) + \text{sign}(w_{i,j}(n) -w_{i,j}(n-1))\Delta w(n)$} \\
    \COMMENT {$\Delta w(n) = l_1 e^{-\frac{n}{l_2}}$}
		\ENDIF
    \ENDFOR
    \ENDWHILE
    \STATE {Send the $i^{th}$ eNodeB' shadow price $p_i^{\text{offered}}$ = $p_i^{\text{offered}}(n) = \frac{\sum_{j=1}^{M_i}w_{i,j}(n)}{R_i}$ to all UEs in the eNodeB coverage area}
\ENDLOOP
    \IF {The $i^{th}$ eNodeB received Flag assignment of $1$ from each UE (the $j^{th}$ UE where $j \in \mathcal{M}_i$) in its coverage area}%
\LOOP
	\STATE{Let $w_i^j(0) = 0\:\:\forall j, j=\{1,....,M_i\}$}
    \WHILE {$|w_i^j(n) -w_i^j(n-1)|>\delta$ for any $j=\{1,....,M_i\}$} %
    \STATE {Calculate $p_i(n) = \frac{\sum_{j=1}^{M_i}w_i^j(n)}{R_i}$}
    \FOR{$j \leftarrow 1$  to  $M_i$}%
    \STATE {Receive $c_i^j$ value from the $j^{th}$ UE}%
    \STATE {Solve $r_i^j(n) = \arg \underset{r_i^j}\max \Big(\log U_j(r_i^j+c_i^j) - p_i(n)r_i^j(n)\Big)$}%
    \STATE {Calculate new $w_i^j(n)= p_i(n) r_i^j(n)$}
    \IF {$|w_i^j(n) - w_i^j(n-1)| > \Delta w$}
    \STATE {$w_i^j(n) =w_i^j(n-1) + \text{sign}(w_i^j(n) -w_i^j(n-1))\Delta w(n)$} \\
    \COMMENT {$\Delta w(n) = l_1 e^{-\frac{n}{l_2}}$}
		\ENDIF
    \ENDFOR	
        \ENDWHILE
    \STATE {Send rate $r_i^{j,opt}= \frac{w_i^j(n)}{R_i}$ to all UEs in the eNodeB coverage area}%
    \STATE {Send the shadow price $p_i=p_i(n)$ to all UEs in its coverage area}%
\ENDLOOP
    \ENDIF
\end{algorithmic}
\end{algorithm}
\section{Simulation Results}\label{sec:sim}
\begin{figure}
\includegraphics[height=1.4in, width=3.5in]{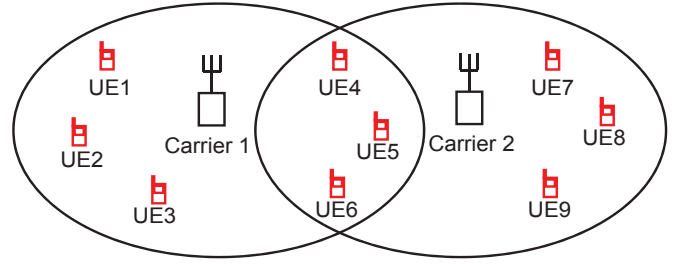}
\caption{System model with two carriers eNodeBs and three groups of users. UE1,UE2 and UE3 under the coverage area of only carrier $1$. UE4, UE5 and UE6 under the coverage area of both carriers. UE7, UE8 and UE9 under the coverage area of only carrier $2$.}
\label{fig:SystemModel}
\end{figure}

\begin{figure}
\includegraphics[height=1.6in, width=3.5in]{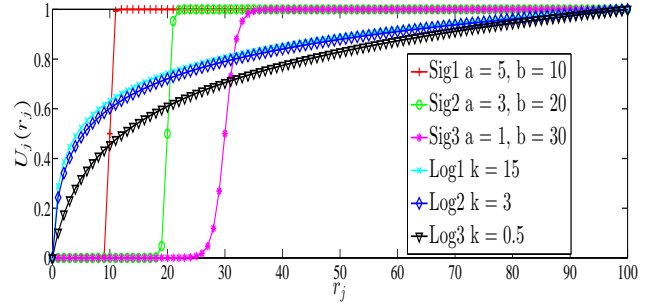}
\caption{The users utility functions $U_j (r_j)$. Sig1 represents UE1 and UE7 applications, Sig2 represents UE2 and UE8 applications, Log1 represents UE3 and UE9 applications, Log2 represents UE4 application, Log3 represents UE5 application and Sig3 represents UE6 application, $r_j$ is the rate allocated to the $j^{th}$ user from all in range eNodeBs.}
\label{fig:utility}
\end{figure}

Algorithm \eqref{alg:UE_FK} and Algorithm \eqref{alg:eNodeB_FK} were applied in C++ to different sigmoidal-like and logarithmic utility functions. The simulation results showed convergence to the global optimal rates. In this section, we present the simulation results for two carriers and $9$ UEs shown in Figure \ref{fig:SystemModel}. Three UEs \{UE1,UE2,UE3\} (first group) are under the coverage area of only Carrier $1$ eNodeB, another three UEs \{UE4,UE5,UE6\} (second group) are joint users under the coverage area of both carrier $1$ and carrier $2$ eNodeBs and three UEs \{UE7,UE8,UE9\} (third group) are under the coverage area of only carrier $2$ eNodeB.
UE1 and UE7 are running the same real-time application that is represented by a normalized sigmoidal-like utility function, that is expressed by equation (\ref{eqn:sigmoid}), with $a = 5$, $b=10$ which is an approximation to a step function at rate $r =10$. UE2 and UE8 are running the same real-time application that is represented by another sigmoidal-like utility function with $a = 3$ and $b=20$. UE3 and UE9 are running the same delay-tolerant application that is represented by a logarithmic function with $k=15$. The joint users UE4 and UE5 are running delay tolerant applications that are represented by logarithmic functions with $k=3$ and $k=0.5$, respectively. The joint user UE6 is running real-time application that is represented by sigmoidal-like utility function with $a=1$ and $b=30$. Additionally, We use $r_{max}=100$ for all logarithmic functions, $l_1=5$ and $l_2=10$ in the fluctuation decay function of the algorithm and $\delta=10^{-3}$. The utility functions corresponding to the nine UEs applications are shown in Figure \ref{fig:utility}.
\subsection{The $i^{th}$ carrier offered Price $p_i^{\text{offered}}$ for $50 \le R_1 \le 200$ and $R_2=100$}
\begin{figure}
\includegraphics[height=1.6in, width=3.5in]{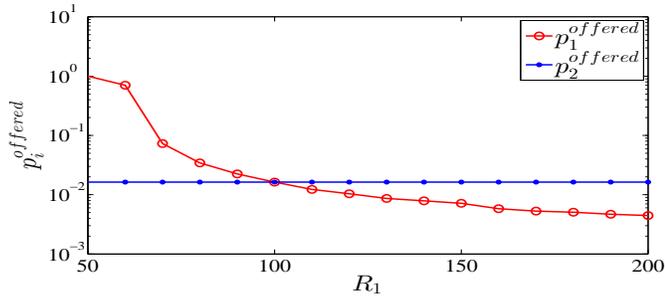}
\caption{Carrier $1$ offered price $p_1^{\text{offered}}$ for different values of $R_1$ and fixed number of users and carrier $2$ offered price $p_2^{\text{offered}}$ for $R_2=100$ assuming that each carrier is the primary carrier for all UEs under its coverage area.}
\label{fig:price_differentRs}
\end{figure}

In the following simulations, carrier $1$ eNodeB available resources $R_1$ takes values between $50$ and $200$ with step of $10$, and carrier $2$ eNodeB available resources is fixed $R_2=100$. In Figure \ref{fig:price_differentRs}, we consider each carrier to be the primary carrier for all UEs under its coverage area and show that carrier $1$ offered price $p_1^{\text{offered}}$ is higher than carrier $2$ offered price $p_2^{\text{offered}}$ when $R_1 \leq R_2$ where $R_2=100$. On the other hand, Figure \ref{fig:price_differentRs} shows that $p_2^{\text{offered}}>p_1^{\text{offered}}$ when $R_2<R_1\le200$. This shows how the carrier's offered price depends on its available resources, the shadow price increases when the carrier's available resources decreases for a fixed number of users. As mentioned before, the joint users select the carrier with the lowest offered price to be their primary carrier. Therefore, in this case the joint users select carrier $2$ to be their primary carrier and carrier $1$ to be their secondary carrier when $R_1 \leq 100$ whereas they select carrier $1$ to be their primary carrier and carrier $2$ to be their secondary carrier when $100<R_1\le200$.
\subsection{Aggregated rates $r_j^{agg}$ for $50 \le R_1 \le 200$ and $R_2=100$}
\begin{figure}
\includegraphics[height=1.6in, width=3.5in]{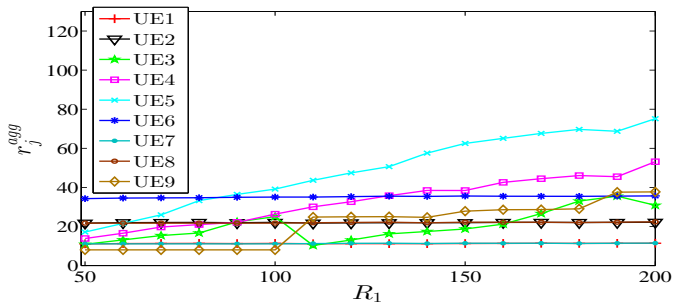}
\caption{The aggregated final optimal allocated rate $r_j^{agg}$ for each user from its all in range carriers versus carrier $1$ available resources $50 \le R_1 \le 200$ with carrier $2$ available resources fixed at $R_2=100$.}
\label{fig:ri_differentRs}
\end{figure}

In the following simulations, carrier $1$ available resources $R_1$ takes values between $50$ and $200$ with step of $10$ and carrier $2$ eNodeB available resources is fixed $R_2=100$.
In Figure \ref{fig:ri_differentRs}, we show the aggregated final optimal rates for the nine users with different available resources  $R_1$ of carrier $1$. The final optimal rates $r_j^{agg}$ for the first group of UEs are allocated to them by only carrier $1$ as they are under the coverage area of only that carrier and do not have secondary carriers. Similarly, the final optimal rates $r_j^{agg}$ for the third group of UEs are allocated to them by carrier $2$ as they are under the coverage area of only that carrier and do not have secondary carriers. On the other hand, the second group of UEs are joint users and are allocated rates from both carriers. The joint users select their primary carrier $l_1^j$ to be the carrier with the lowest shadow price $l_1^j=\arg \underset{\{1,2\}} \min \{p_1^{\text{offered}},p_2^{\text{offered}}\}$ and the other carrier with a higher offered price to be their secondary carrier $l_2^j$. The aggregated final optimal rate allocated to each joint user is the aggregated rate of its primary carrier allocated rate and its secondary carrier allocated rate. Figure \ref{fig:ri_differentRs} shows that users running real-time applications are given priority over users running delay tolerant applications and are allocated higher rates in the case of low carrier's available resources. 
\section{Summary and Conclusions}\label{sec:conclude}
In this paper, we introduced a novel RA with CA optimization problem in cellular networks. We considered mobile users with elastic or inelastic traffic and used utility functions to represent the applications running on the UEs. We presented a novel price selective centralized algorithm for allocating resources from different carriers optimally among users. Our price selective algorithm guarantees the minimum possible price for the aggregated final optimal rates. The algorithm uses proportional fairness approach to provide a minimum QoS to all users while giving priority to real-time application users. Our centralized algorithm is performed mostly in the eNodeBs. Therefore, it requires less transmission overhead and less computations in the UEs. We analyzed the convergence of the algorithm with different carriers available resources and showed through simulations that our algorithm converges to optimal values.
\bibliographystyle{ieeetr}
\bibliography{pubs}

\begin{thebibliography}{10}

\bibitem{QoScontrol}
H.~Ekstrom, ``Qos control in the 3gpp evolved packet system,'' {\em
  Communications Magazine, IEEE}, vol.~47, no.~2, pp.~76--83, 2009.

\bibitem{UtilityMax}
Y.~Wang, K.~Pedersen, T.~Sorensen, and P.~Mogensen, ``{Utility Maximization in
  LTE-Advanced Systems with Carrier Aggregation},'' in {\em Vehicular
  Technology Conference (VTC Spring), 2011 IEEE 73rd}, pp.~1--5, May 2011.

\bibitem{Yuan10CarrierAggregation}
G.~Yuan, X.~Zhang, W.~Wang, and Y.~Yang, ``{Carrier aggregation for
  LTE-advanced mobile communication systems},'' in {\em Communications
  Magazine, IEEE}, vol.~48, pp.~88--93, 2010.

\bibitem{Evolution}
S.~Parkvall, A.~Furuskar, and E.~Dahlman, ``{Evolution of LTE toward
  IMT-advanced},'' {\em Communications Magazine, IEEE}, vol.~49, pp.~84--91,
  February 2011.

\bibitem{Haya_Utility1}
H.~Shajaiah, A.~Abdel-Hadi, and C.~Clancy, ``Utility proportional fairness
  resource allocation with carrier aggregation in 4g-lte,'' in {\em Military
  Communications Conference, MILCOM 2013 - 2013 IEEE}, pp.~412--417, Nov 2013.

\bibitem{kelly98ratecontrol}
F.~Kelly, A.~Maulloo, and D.~Tan, ``Rate control in communication networks:
  shadow prices, proportional fairness and stability,'' in {\em Journal of the
  Operational Research Society}, vol.~49, 1998.

\bibitem{Fundamental}
S.~Shenker, ``Fundamental design issues for the future internet,'' {\em
  Selected Areas in Communications, IEEE Journal on}, vol.~13, no.~7,
  pp.~1176--1188, 1995.

\bibitem{RebeccaThesis}
R.~Kurrle and C.~Clancy, ``{Resource Allocation for Smart Phones in 4G-LTE
  Advanced Carrier Aggregation},'' Master's thesis, Virginia Polytechnic
  Institute and State University, 2012.

\bibitem{Ahmed_Utility1}
A.~Abdel-Hadi and C.~Clancy, ``{A Utility Proportional Fairness Approach for
  Resource Allocation in 4G-LTE},'' in {\em ICNC Workshop CNC}, 2014.

\bibitem{Ahmed_Utility2}
A.~Abdel-Hadi and C.~Clancy, ``{A Robust Optimal Rate Allocation Algorithm and
  Pricing Policy for Hybrid Traffic in 4G-LTE},'' in {\em PIMRC}, 2013.

\bibitem{Ahmed_Utility3}
A.~Abdel-Hadi, C.~Clancy, and J.~Mitola, ``{A Resource Allocation Algorithm for
  Multi-Application Users in 4G-LTE},'' in {\em MobiCom Workshop}, 2013.

\bibitem{Haya_Utility2}
H.~Shajaiah, A.~Abdel-Hadi, and C.~Clancy, ``Spectrum sharing between public
  safety and commercial users in 4g-lte,'' in {\em Computing, Networking and
  Communications (ICNC), 2014 International Conference on}, pp.~674--679, Feb
  2014.

\bibitem{Haya_Utility4}
H.~Shajaiah, A.~Abdel-Hadi, and C.~Clancy, ``{Multi-Application Resource
  Allocation with Users Discrimination in Cellular Networks},'' in {\em
  Accepted in PIMRC, 2014}, 2014.

\bibitem{Haya_Utility3}
H.~Shajaiah, A.~Khawar, A.~Abdel-Hadi, and T.~Clancy, ``Resource allocation
  with carrier aggregation in lte advanced cellular system sharing spectrum
  with s-band radar,'' in {\em Dynamic Spectrum Access Networks (DYSPAN), 2014
  IEEE International Symposium on}, pp.~34--37, April 2014.

\bibitem{DL_PowerAllocation}
J.-W. Lee, R.~R. Mazumdar, and N.~B. Shroff, ``Downlink power allocation for
  multi-class wireless systems,'' {\em IEEE/ACM Trans. Netw.}, vol.~13,
  pp.~854--867, Aug. 2005.

\bibitem{Utility-proportional}
G.~Tychogiorgos, A.~Gkelias, and K.~K. Leung, ``Utility-proportional fairness
  in wireless networks,'' in {\em Personal Indoor and Mobile Radio
  Communications (PIMRC), 2012 IEEE 23rd International Symposium on},
  pp.~839--844, Sept 2012.

\end{thebibliography}
\end{document}